\documentclass[fleqn,usenatbib]{mnras}

\usepackage{newtxtext,newtxmath}
\usepackage[T1]{fontenc}
\usepackage{ae,aecompl}
\usepackage{graphicx}
\usepackage{amsmath}
\usepackage{lipsum}

\usepackage{multirow}
\usepackage[section]{placeins}
\usepackage{mathrsfs}
\usepackage{ulem}
\usepackage[dvipsnames]{xcolor}
\usepackage[caption=false]{subfig}

\newcommand\lsim{\mathrel{\rlap{\lower4pt\hbox{\hskip1pt$\sim$}}
        \raise1pt\hbox{$<$}}}
\newcommand\gsim{\mathrel{\rlap{\lower4pt\hbox{\hskip1pt$\sim$}}
        \raise1pt\hbox{$>$}}}

\title[Periodic Quasars in LSST as future LISA sources]{Ultra-Short-Period Massive Black Hole Binary Candidates in LSST as LISA "Verification Binaries"}

\author[Xin \& Haiman]{Chengcheng~Xin,$^{1}$
Zolt{\'{a}}n~Haiman,$^{1}$ \\
$^{1}$Department of Astronomy, Columbia University, New York, NY, 10027, USA\\}

\date{Accepted XXX. Received YYY; in original form ZZZ}

\pubyear{2020}

\begin{document}
\label{firstpage}
\pagerange{\pageref{firstpage}--\pageref{lastpage}}
\maketitle

\begin{abstract}
The Legacy Survey of Space and Time (LSST) by the Vera C. Rubin Observatory is expected to discover tens of millions
of quasars.
A significant fraction of these could be powered by coalescing massive black hole (MBH) binaries, since many quasars are believed to be triggered by mergers.
We show that under plausible assumptions about the luminosity functions, lifetimes, and binary fractions of quasars, we expect the full LSST quasar catalogue to contain 
between 20-100 million compact MBH binaries with masses $M=10^{5-9}M_{\odot}$, redshifts $z=0-6$, and orbital periods $P=1-70$ days. 
Their light-curves are expected to be distinctly periodic, which can be confidently distinguished from stochastic red-noise variability, because LSST will cover dozens, or even hundreds of cycles.
A very small subset of 10-150 ultra-compact ($P\lsim 1$ day) 
binary quasars among these will, over $\sim$5-15 years, evolve into
the mHz gravitational-wave (GW) frequency band and can be detected by {\it LISA}.  They can therefore be regarded as "{\it LISA} verification binaries", analogous to short-period Galactic compact-object binaries.   The practical question is how to find these handful of "needles in the haystack" among the large number of quasars: this will likely require a tailored co-adding analysis optimised for this purpose.
\end{abstract}


\section{Introduction}\label{sec:intro}

The Laser Interferometer Space Antenna ({\it LISA}; \citealt{LISA}) is currently planned for launch in the mid 2030s and is expected to discover gravitational waves (GWs) from coalescing massive black hole (MBH) binaries with component masses $\approx 10^4-10^7~{\rm M_\odot}$ out to high redshifts.  It has been widely recognised that detecting the electromagnetic (EM) emission from these binaries, in addition to their GWs, will enable a significant range of novel science investigations, ranging from a better understanding accretion onto BHs, to the connection between MBHs and their host galaxies, and to new tests of general relativity~\citep{MMMBA-decadal}.

The EM emission from a {\it LISA} source could be searched for in follow-up observations to {\it LISA} events, or possibly by EM observations during the late inspiral stage, triggered by a {\it LISA} alert~\citep{Kocsis+2008}.
This is different from the situation for Galactic white-dwarf binaries, for which several are now known that will be detectable by {\it LISA} - the so-called "Verification Binaries" ~\citep[see][and references therein]{Burdge+2020}.

Here we suggest that MBH binaries detectable by {\it LISA} may also be identified prior to {\it LISA}'s launch, among the large number of active galactic nuclei (AGN) expected to be catalogued by the Vera C Rubin Observatory's Legacy Survey of Space and Time (LSST).   Since most major galaxies contain nuclear MBHs~\citep{KormendyHo2013}, and in hierarchical structure formation, galaxies are built up via mergers, pairs of MBHs  should be commonly found in galactic nuclei.  MBHs are expected to rapidly sink to the nucleus of the new galaxy and form a gravitationally bound binary~\citep{Begelman+1980}.   Furthermore, torques  during galaxy mergers are understood to drive gas to the nucleus of the new galaxy~\citep{BarnesHernquist1991}, and
bright and relatively short-lived quasar phases can be triggered by mergers~\citep{KauffmannHaehnelt2000,AlexanderHickox2012}. The association between galaxy mergers and AGN activity is also supported by observation~\citep{Goulding+2018}.    

Over the past decade,  hydrodynamical simulations of binary BHs with circumbinary discs have reached the consensus that a binary can continue accreting efficiently from the disc, via narrow, collimated gas streams, essentially at the same rate as a single MBH would~\citep{Dorazio+2013,Farris+2014,ShiKrolik2015,Miranda+2017}.  Moreover, efficient accretion has been shown to persist all the way to merger~\citep{Farris+2015,Tang+2018,Bowen18,Bowen19}.  The light-curves of MBHBs can therefore be expected to be as bright as quasars, but should contain significant periodicities, due either to hydrodynamical modulations of the accretion rate (as shown in the above simulations), and/or due to relativistic Doppler and lensing modulations of the apparent flux~\citep{Bogdanovic+2011,Dorazio+2015,Haiman2017,DorazioDiStefano2018}.  As a result, MBH binaries may be identifiable in time-domain AGN surveys as a population of periodic quasars~~\citep{Haiman2009a}.

In recent years, significant effort has indeed been invested to find periodic quasars in large time-domain quasar catalogues, e.g. in the Catalina Real-time Transient Survey (CRTS; \citealt{Graham+2015}), the Palomar Transient Factory (PTF; \citealt{Charisi+2016}, and in a joint analysis of the Dark Energy Survey and the Sloan Digital Sky Survey~(DES+SDSS; \citealt{Chen+2020}.  (See the recent review by \citealt{DeRosa+2019} for a more complete summary.)

These efforts have identified $\approx200$ periodic candidates among $O(10^5)$ quasars, with BH masses of $10^{8-9}~{\rm M_\odot}$ and putative periods ranging from several months to a few years.  This roughly matches expectations: massive binary BHs are expected to spend $O(10^5)$ years at periodicities of order a year, if their orbital decay is driven either by GW emission or by negative torques exerted by the surrounding gas disc on the viscous timescale; \citealt{Haiman2009a}). If quasars typically live for a few  $\times 10^7$ years~\citep{MartiniSchneider2004}, then we might expect a fraction of few $\times 10^{-3}$ to be in this stage.

The LSST is expected to deliver a revolutionary large sample of at least tens of millions of quasars, each with well-sampled light-curves at high cadence (i.e every few days). These two novel characteristics will allow a search for MBH binaries with much shorter-periods, which are correspondingly much rarer.

In this paper, we suggest that it may be possible to identify ultra-short-period MBH binaries in the LSST quasar catalogue, which are so compact that they will "chirp" -- i.e. evolve in frequency -- into the mHz GW band, where {\it LISA} can subsequently detect them several years later.   If so, these sources would be the massive BH analogues of the Galactic "Verification Binaries".  We will argue that this is possible under plausible assumptions
about the luminosity functions, lifetimes, and binary fractions of quasars.    In particular, if (i) the faint of end of the quasar luminosity function is steep and can be extrapolated down luminosities corresponding to BH masses of $\sim 10^5~{\rm M_\odot}$, (ii)  most quasars are associated with merging MBH binaries, and (iii) quasar lifetimes are relatively short ($\lsim 10^8$ years), then we expect the LSST's quasar catalogue to contain several such ultra-compact MBH binaries with periods $\lsim$ day.
Most of these "Verification Binaries" will lie below LSST's single-snapshot flux detection threshold, and identifying them will likely require co-adding individual snapshots in an optimised way.

The rest of this paper is organised as follows. 
In \S~\ref{sec:method}, we discuss our basic methodology, which consists of adopting a quasar luminosity function (\S~\ref{sec:QLF}), computing the corresponding LSST quasar number counts (\S~\ref{sec:N_tot}), and considering detectability by {\it LISA} (\S~\ref{sec:lisa_sensitivity}.
In \S~\ref{sec:results}, we present our main results, i.e. the expected number of {\it LISA}-detectable MBH binaries among the LSST quasars, as well as how these results depend on the assumed quasar lifetime and typical binary mass ratio.
In \S~\ref{sec:discussion}, we further discuss several caveats and implications of our results, including the requirements for co-adding LSST data and folding quasar light-curves.
Finally, our conclusions are summarised in \S~\ref{sec:conclusion}.

\section{Method} \label{sec:method}

The efforts in large surveys of active galactic nuclei (AGN), including bright quasars, have been growing remarkably in the last few decades. This is reflected not only in the significant expansion of quasar samples, up to a size of a $\sim$ million sources~\citep{Flesch2019},
but also in the depth of individual surveys (see compilation in \citealt{Kulkarni2019}). The upcoming LSST will offer a major improvement in the observations of quasars in survey area and depth, as well as in variability information, with the light-curves of most sources sampled at high cadence (of order days) over several years~\citep{LSSTScienceCollaboration2009,Ivezic2019}.
In this section, we introduce the quasar luminosity functions we use to approximate the number of quasars that LSST is expected discover. 

A fraction of variable quasars in LSST might be compact massive black hole binaries (MBHBs), produced in mergers which triggered the quasar activity~\citep{KauffmannHaehnelt2000,Haiman2009a}
and emitting gravitational waves (GWs) prior to their coalescence. A small subset of these objects, i.e. those with ultra-short (sub-day) periods, will evolve into the mHz GW band, where they can be discovered by {\it LISA} after accumulating sufficient signal-to-noise ratio.
To find potential {\it LISA} detections of ultra-short--period quasars in LSST, we first estimate the number of binary quasars in LSST, and then employ the most up-to-date {\it LISA} sensitivity curve from \citealt{Robson2019}. We record a complete set of acceptable orbital periods within the {\it LISA} frequency band. Finally, we compute the number of binaries detectable by {\it LISA} that is obtained from a simple relationship between the binary merger time and the total quasar lifetime. 

\subsection{Quasar Luminosity Functions} \label{sec:QLF}

The number of luminous quasars is given directly by the quasar luminosity function (QLF). Quasars are considered to be produced by accretion onto massive black holes at the centres of galaxies \citep{Lynden1969}, and their activity is thought to be often triggered by the merger of two galaxies, delivering fuel to the nucleus, activating accretion \citep[e.g.][]{BarnesHernquist1991}
Quasar LFs can therefore be used to determine the abundance of MBHs at different cosmological redshifts. 
Since both merging galaxies triggering the quasar activity are believed to typically contain nuclear MBHs, a large fraction of quasars could contain binary MBHs~\citep[e.g.][]{Haiman2009a,Dotti+2015}.  

Over the last few decades, a large amount of work has gone into understanding the shape and evolution of the QLFs throughout cosmic history. In this paper, we have chosen to use the QLFs reported in \citet[][hereafter K19]{Kulkarni2019}, for several reasons. K19 is based on a recent comprehensive compilation, implementing QLFs from a large homogenised AGN sample across a wide range of redshifts, up to z$\sim$7. Their LFs are also conveniently given in the form of simple broken power-laws  as a function of absolute magnitude, with the band-pass conversions from {\it apparent} optical magnitudes to absolute magnitude explicitly provided (i.e. their Fig.2). This latter feature is especially convenient,  since it allows us to directly use them to compute the number of quasars expected to be observed by LSST above a given apparent magnitude threshold.\footnote{This is unlike the case for the large majority of other QLF determinations in the literature~\citep[e.g.][and references therein]{Shen+2020}, which quote QLFs only as a function of absolute magnitude or luminosity, which would make it necessary for us to "undo" the band flux $\rightarrow$ luminosity computations in those works.}

\begin{figure}
    \centering
    \includegraphics[width=\columnwidth]{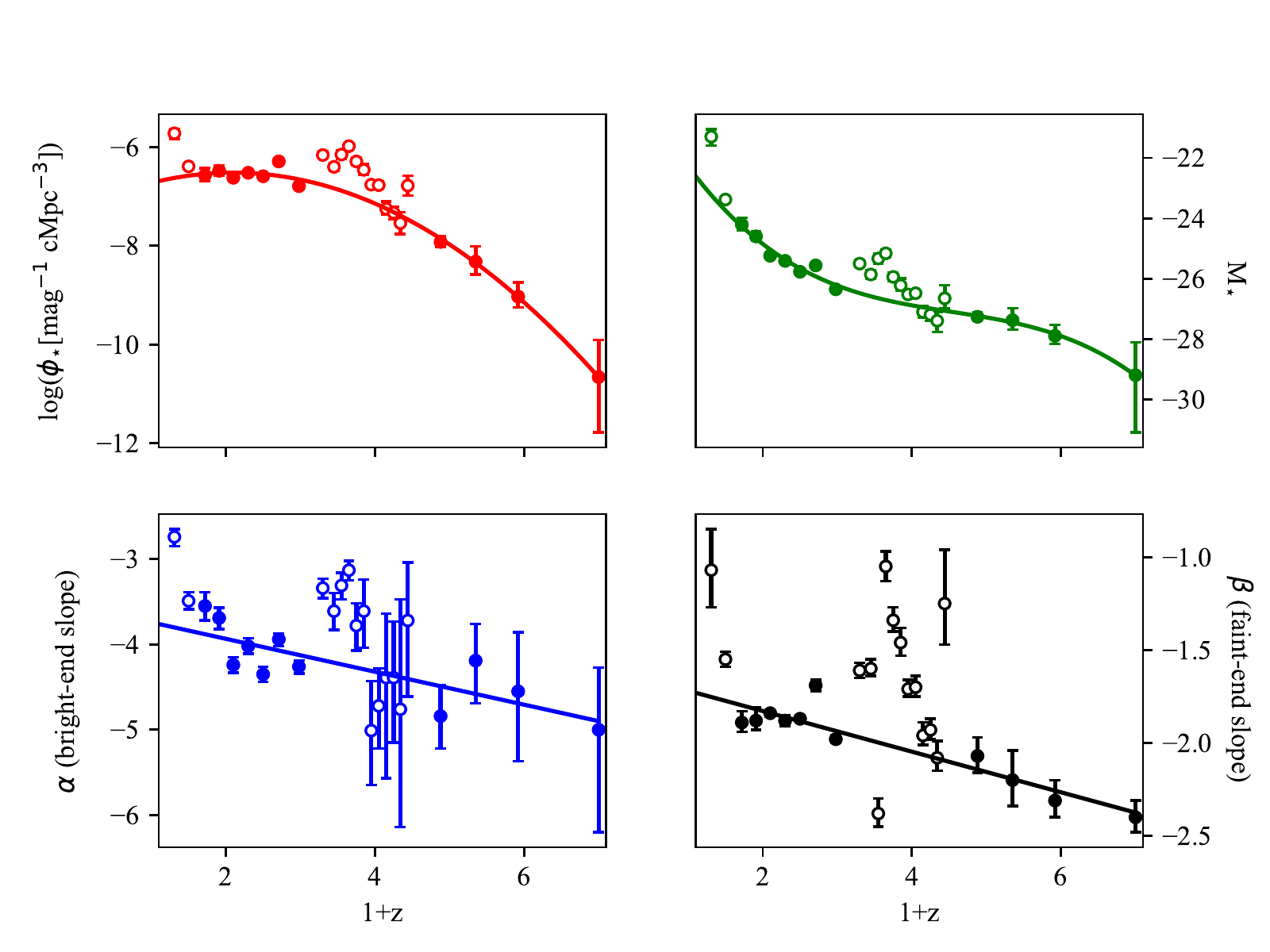}
    \caption{Double power--law parameters in the quasar luminosity function as a function of redshift (clockwise from upper left: normalisation, characteristic magnitude, faint end-slope, and bright-end slope). We show the parameter values at the 25 effective redshifts in K19. We fit the solid circles, which are retained in the fiducial model in K19, with polynomials (top panels) and linear curves (bottom panels). The open circles that appear discrepant from the solid points are subject to large systematic uncertainties in the AGN data from different surveys, and are therefore discarded from the fit. }
    \label{fig:parameters}
\end{figure}

As mentioned above, K19 use a double power--law (PL) to fit the space density of quasars versus apparent magnitude. There are four parameters that define the double PL, which vary as functions of redshift, 
\begin{equation} \label{eq:qlf}
    \phi(M)=\frac{\phi^{\star}}{10^{0.4(\alpha+1)(M-M_{\star})}+10^{0.4(\beta+1)(M-M_{\star})}}.
\end{equation}
The parameters are the overall normalisation $\phi^{\star}$ given in units cMpc$^{-3}$mag$^{-1}$, the break magnitude $M_{\star}$ and the bright-- and faint--end slopes, $\alpha$ and $\beta$.
The fiducial values of these parameters, as reported in K19 at 25 different redshifts, are shown in Figure~\ref{fig:parameters}. The normalisation $\phi_{\star}$ is roughly constant with redshift
between $0\lsim z\lsim 3$ and then decreases sharply, by $\sim$3 orders of magnitude by $z\approx 5$. The break magnitude $M_{\star}$ decreases monotonically between $0\lsim z\lsim 7$ . The bright--end ($\alpha$) and the faint--end ($\beta$) slopes both show a general trend to steepen with redshift, with visible discrepancies around $z\sim$2-3, shown by empty circles.  As explained in K19, compared with the relatively smooth evolution of other parameters, the sharp steepening of $\beta$ at $z\sim 2.5$ and dramatic flattening near this redshift are likely unphysical and instead indicate a mismatch between the BOSS and SDSS+2SLAQ surveys due to systematic errors induced by their selection functions. Likewise, the outliers at low redshift ($z<0.6$; also shown as empty circles) result from uncertainties in the correction for host galaxy light and potentially missed AGN in extended sources. As a result, we follow K19, who excluded these points from fitting parametric redshift-dependent QLFs,  and also neglect these points (i.e. all open circles) 
in our fiducial analysis.
Instead, we interpolate the remaining reliable points (solid circles Fig.~\ref{fig:parameters}), and obtain values of each parameter as smooth functions of redshift. Specifically, we fit 2$^{\rm nd}$ and 3$^{\rm rd}$ order polynomials to the solid circles in the top panels from left to right and straight lines in the bottom panels.  We find that using higher-order polynomials is not warranted.
 
\begin{figure}
  \centering
    \includegraphics[width=1.1\columnwidth]{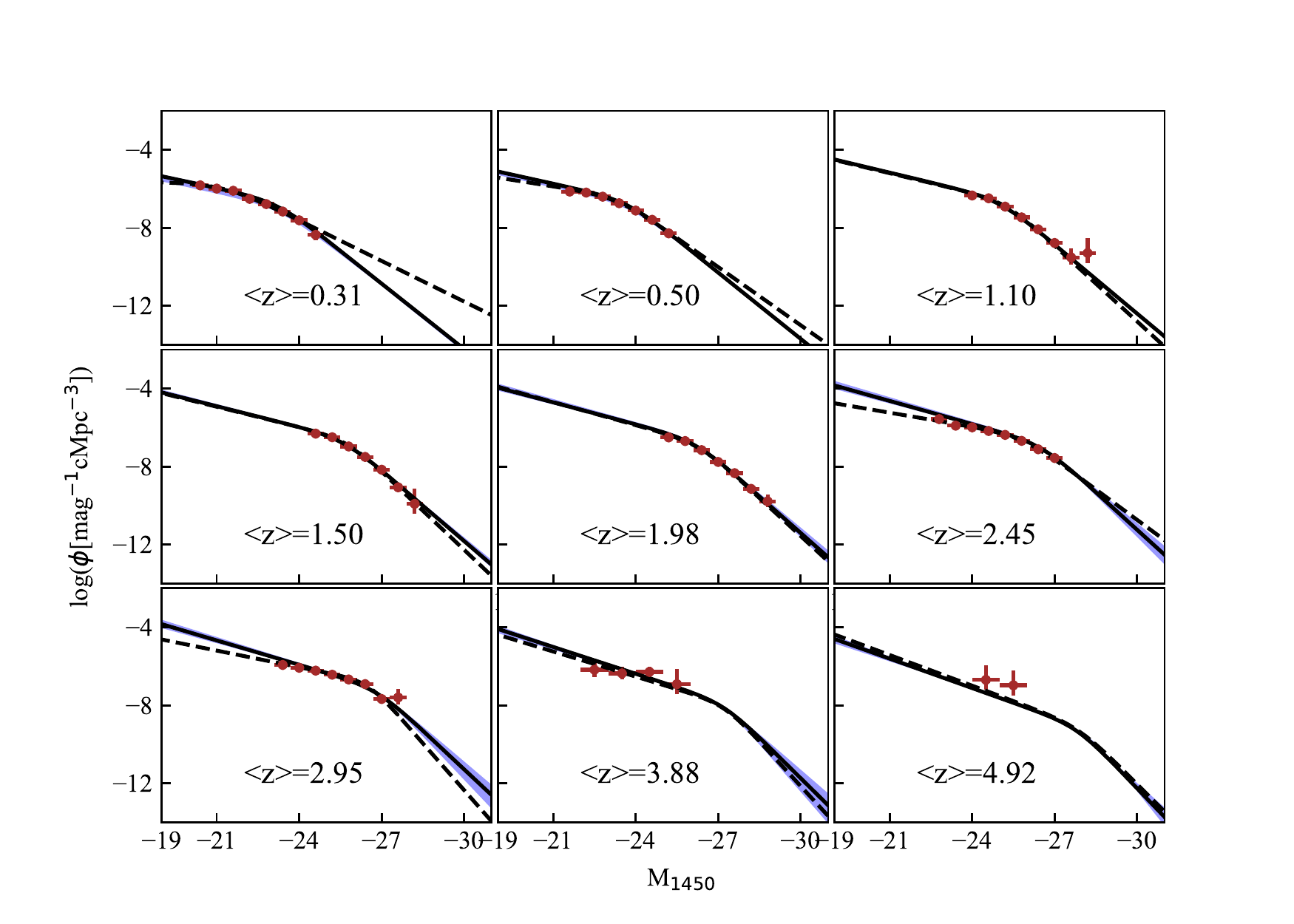}
    \caption{Double power--law quasar luminosity functions (LFs) in 9 different redshift bins with central values as indicated in each panel. The fiducial results from K19 are reproduced by the dashed curves. Our LFs with the four parameters interpolated smoothly across different redshifts are shown by the solid curves. The red circles and their error bars represent the binned AGN data used in K19. The regions in shade blue indicate the upper and lower uncertainties in LFs obtained from interpolating the error bars in Figure~\ref{fig:parameters}.}
    \label{fig:LF-compare}
\end{figure}

We demonstrate the QLFs obtained from these fits in 9 randomly selected redshift bins between $0.1 < z < 6$ with bins centred at $z=\{0.31, 0.50, 1.10, 1.50, 1.98, 2.45, 2.95, 3.88, 4.92\}$ and widths $\Delta z\approx $0.2-0.4. These redshift bins are adopted from Table~2 in K19 and can be used to compare our QLFs to theirs, as shown in Figure~\ref{fig:LF-compare}. We compare the double PLs with parameters found in K19 (dashed) and those with our newly interpolated parameters (solid) as functions of absolute UV magnitude (at rest-frame  $\lambda$1450\AA). We also colour the uncertainty regions in each panel in blue, using the interpolated upper and lower errors shown in Figure~\ref{fig:parameters}.  While K19 base their analysis on a large compilation of AGN samples from different surveys, each sample is categorised into smaller magnitude bin(s). We superimpose the binned AGN data in red, where the shape of the LF is directly measured (and thus more reliable) than at either fainter or brighter magnitudes, where they are extrapolated. 
The interpolated curves show a larger degree of continuity in both the bright-- and the faint--end slopes across different redshift. 

\begin{figure*}
    \centering
    \includegraphics[width=2\columnwidth]{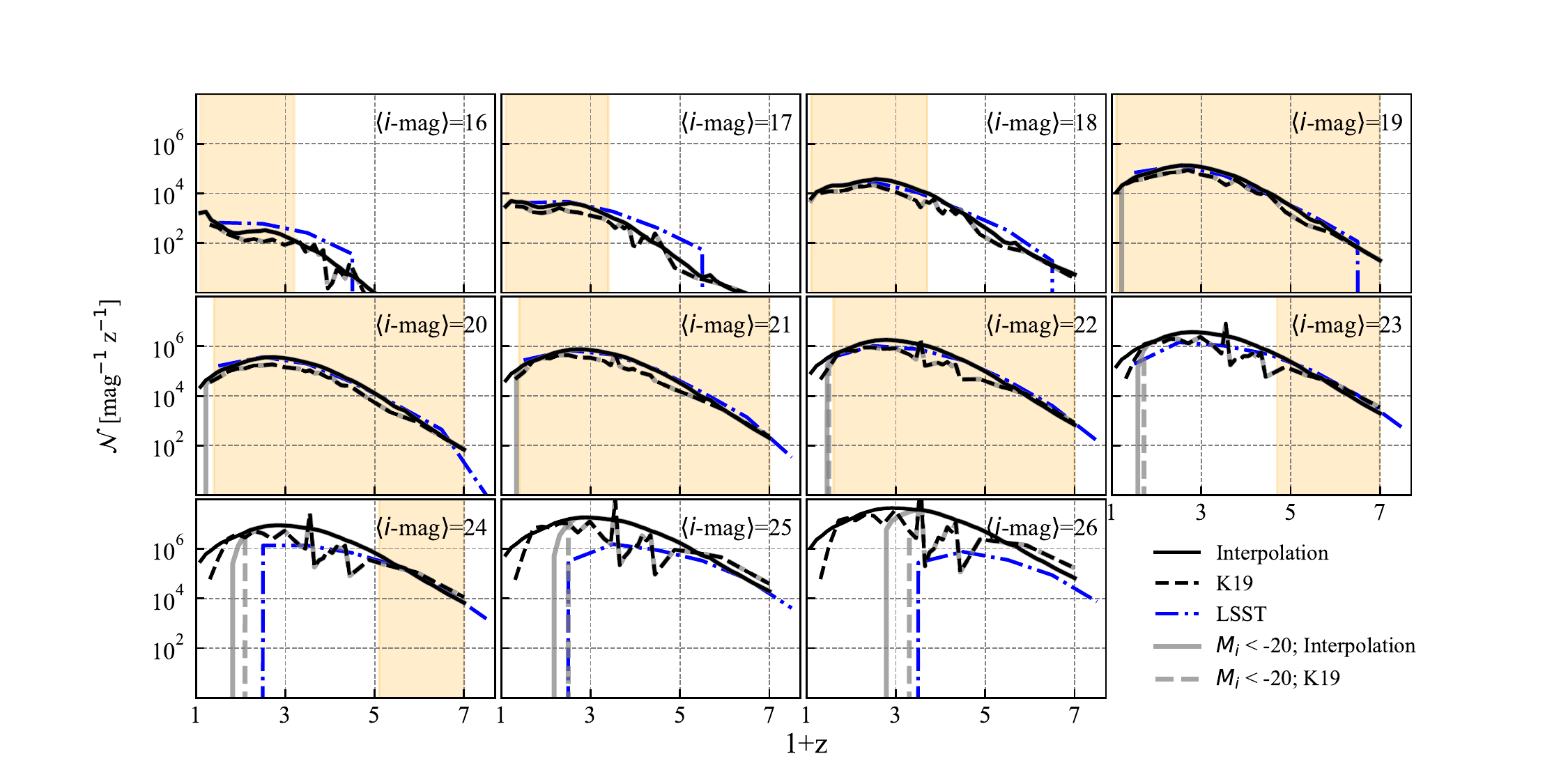}
    \caption{The total number of quasars expected in LSST as a function of redshift, in 11 different apparent $i$-magnitude bins. $\mathcal{N}$ has units of ${\rm mag}^{-1} z^{-1}$, 
    and is shown as a function of $(1+z)$ in different $i$ magnitude bins, as indicated in each panel. The different curves correspond to our interpolated LFs (solid black), K19's fiducial LFs (dashed black), the counts in the LSST Science Book (dot-dashed blue),
    and both our and K19's predictions with an additional cut $M_i<-20$ imposed on the absolute $i$ magnitude, as done in the LSST Science Book (grey solid and grey dashed, respectively). The vertical light orange bands mark the redshift ranges covered by the quasar sample data in K19 in each magnitude bin; counts shown outside these bands represent extrapolations.}
    \label{fig:N-z-coarse}
\end{figure*}
\subsection{The Number of Short--Period Quasars in LSST} \label{sec:N_tot}
As mentioned above, the Rubin Observatory's LSST is expected to discover a large number of quasars. An early estimate for these numbers which fold in the expected completeness as a function of redshift and apparent magnitude indicate that $\sim 20$ million quasars in total can be identified in single-exposure images \citep{LSSTScienceCollaboration2009}.  This number is estimated by integrating the quasar LF (e.g.~Eq.~\ref{eq:qlf}), over the desired magnitude and redshift intervals. The predictions in \citet{LSSTScienceCollaboration2009} were based on the earlier QLF determinations by \citet{Hopkins2007}.

Here we re-compute the number of quasars in a magnitude and redshift range,
$\mathcal{N}=\mathcal{N}(m_{\rm min},m_{\rm max}; z_{\rm min}, z_{\rm max})$. We convert the absolute UV magnitude ($M_{1450}$) to apparent i--band magnitude, allowing us to relate the mass of the black holes to their observed abundance. We adopt the bandpass $\mathcal{K}$-corrections from K19 (who, in turn, adopt it from \citealt{Lusso2015}, shown in their Fig.~2), which conveniently directly converts $M_{1450}$ to apparent $i$-mag. $\mathcal{N}(m_{\rm min},m_{\rm max}; z_{\rm min}, z_{\rm max})$ is then given by
\begin{equation} \label{eq:V_eff}
    \mathcal{N} = \int_{m_{\rm min}}^{m_{\rm max}} dm \int_{z_{\rm min}}^{z_{\rm max}} dz f(m,z)\phi(M_{1450}(m),z) \frac{dV}{dz},
\end{equation}
where $f(m,z)$ is the selection function characterising the completeness, i.e. the fraction of selected AGNs to the true number of AGNs in a certain volume and magnitude range. The selection function needs to account for all of the ways that AGNs are identified in LSST, i.e. via colours, the lack of proper motion, variability and combinations with other multi-wavelength surveys~\citep{LSSTScienceCollaboration2009}. In general, this makes selection complex, with $f$ a function of the magnitude and redshift ranging from 0 to 1. In this analysis, we approximate $f(M,z)$ as a step function, with $f=0$ at magnitudes above (fainter) than a threshold $m_i$ and $f=1$ for brighter quasars.
The comoving volume element $dV/dz$ is given by
\begin{equation} \label{eq:comov_V}
    \frac{dV}{dz} = \frac{dV}{dzd\Omega}\times A \times \frac{4\pi}{41253},
\end{equation}
where $A$ is the LSST survey area, assumed to be 20,000 deg$^2$ and $dV/dzd\Omega$ is the comoving volume element per solid angle.
We adopt the cosmological parameters $H_0=70~{\rm km~s^{-1}Mpc^{-1}}$, $\Omega_{\Lambda}=0.7$ and $\Omega_{m}=0.3$~\citep{Planck2018}.
\begin{equation} \label{eq:comov_V_sol}
    \frac{dV}{dzd\Omega} = \frac{c}{H_0}\frac{d^2_{L}(z)}{(1+z)^2[\Omega_m(1+z)^3+\Omega_{\Lambda}]^{1/2}}
\end{equation}

Integrating the LF between redshifts $0\leq z\leq 6$ and i--magnitudes $16\leq m\leq 24$ (i.e. adopting a threshold $m_i=24$), we find $\mathcal{N}=20$ million quasars in total.  Extending the magnitude limit it $m_i\leq 26$, the number increases to $100$ million.
 
In Figure~\ref{fig:N-z-coarse}, we show our prediction of $\mathcal{N}$ in 11 different $i$-magnitude bins (obtained numerically be evaluating $\mathcal{N}$ at 50 evenly separated values of $z$ for each $m_i$). For reference, we again show a comparison with predictions based on K19's original QLF determinations (dashed curves) and also the number counts from LSST's Science Book (blue dot-dashed curves). The predictions generally agree with those by LSST, especially at the redshifts most relevant to possible {\it LISA} detections ($z\lsim$2-3; see below).  The vertical shaded orange bands in this figure shows the redshift ranges covered by the AGN samples used to fit the QLFs; redshifts outside these bands represent extrapolations.

For the rest of the analysis, we assume that the majority of the quasars are associated with massive black hole binary mergers. More generally, the number of binaries throughout this paper simply scales linearly with the fraction of quasars $f_{\rm Q,b}$ associated with coalescing black holes. We further assume that the last stages of the MBH binary merger falls within the bright quasar phase.   This is justified by the the expected availability of gas in the nuclei of merger-remnant galaxies, and by recent hydrodynamical simulations of inspiralling MBHs with circumbinary discs~\citep{Farris+2015,Tang+2018,Bowen18}.  These studies found that while the coalescing binary creates a low-density central cavity in the accretion disc, the MBHs can nevertheless be fuelled efficiently, all the way to their merger, via mini-discs fed by narrow accretion streams.  Finally, we assume that
quasars typically have a total lifetime of $t_{\rm Q} \sim {\rm few} \times 10^7$ years (independent of redshift and luminosity), and that they spend a duration of $t(P)$ at orbital period $P$.   Under these assumptions, the number of quasars with orbital periods $P$ is given approximately by $\mathcal{N}(P)\approx \left[t(P) / t_{\rm Q}\right] f_{\rm Q,b} \mathcal{N}$.

The time $t(P)$ that a MBH binary spends at period $P$, for the compact binaries of interest for {\it LISA}, is determined by its GW emission, and depends on the BH masses. To translate the apparent $i$-magnitude into the total mass of binary BHs, we need to assume a quasar luminosity - BH mass relation.  Note that the masses relevant for {\it LISA} are between $\sim 10^5$ and $\sim 10^9$ $M_{\odot}$, with lower-mass binaries dominating the majority of the population~\citep[e.g.][and references therein]{Klein+2016}. 
We adopt the relationship between $i$--magnitude and total BH mass $M_{\rm bh}$ based on a typical quasar spectral energy distribution, and a bolometric quasar luminosity $L=f_{\rm Edd} L_{\rm Edd}$, where $L_{\rm Edd}$ is the Eddington luminosity for mass $M_{\rm bh}$~\citep{Haiman2009a},
\begin{equation}  \label{eq:M-imag}
    m_i = 24 + 2.5 \: {\rm log} \: \bigg[ \bigg( \frac{{\rm f}_{\rm Edd}}{0.3}\bigg) \: \bigg( \frac{M_{\rm bh}}{3\times 10^6 {\rm M}_{\odot}}\bigg)^{-1} \bigg( \frac{d_L(z)}{d_L(z=2)}\bigg)^2 \: \bigg],
\end{equation}
where $d_L(z)$ is the luminosity distance to redshift $z$. With this definition, we compute the distribution of the number of massive BBHs in 50 mass bins evenly distributed across $10^{5-9}M_{\odot}$ in log-space, and 50 redshift bins evenly distributed over $0<z<6$. Therefore our  MBH binary counts are tabulated on a 50$\times$50 grid in the $M_{\rm bh}$--$z$ plane. 

\subsection{Detection by {\it LISA}} \label{sec:lisa_sensitivity}
To decide which MBH binaries can be detectable by {\it LISA}, we use
the recent {\it LISA} sensitivity curve reported in \citet{Robson2019}, which gives the total instrumental noise,  as well as the stochastic noise due to the presence of unresolved Galactic binaries. 
Detectability by {\it LISA} essentially requires that the MBH binary, identified in LSST as an ultra-short-periodic quasar, evolves into {\it LISA}'s frequency band during the time when {\it LISA} is operational.  At the short periods of $O$(days) relevant for our discussion, the binary inspiral is expected to be dominated by GW emission (see below).  In general, it will take at least several weeks for LSST to identify periodicity with $P=O$(days) (since $N\gsim 20$ periods are required; see, e.g. \citealt{Vaughan2016}), and possibly longer if co-adding data from multiple LSST visits and phase-folding of the light-curve is needed.  LSST is currently expected to start survey operations in the mid 2020's and {\it LISA} is currently expected to be launched in the mid 2030's, with a total mission lifetime of 5 years, i.e. ending operations in the early 2040's.   Our requirement is therefore approximately that the time elapsed between the discovery of a periodic source by LSST, and the time it enters the {\it LISA} band, should be between 5-15 years.

\begin{figure}
    \centering
    \includegraphics[width=1.1\columnwidth]{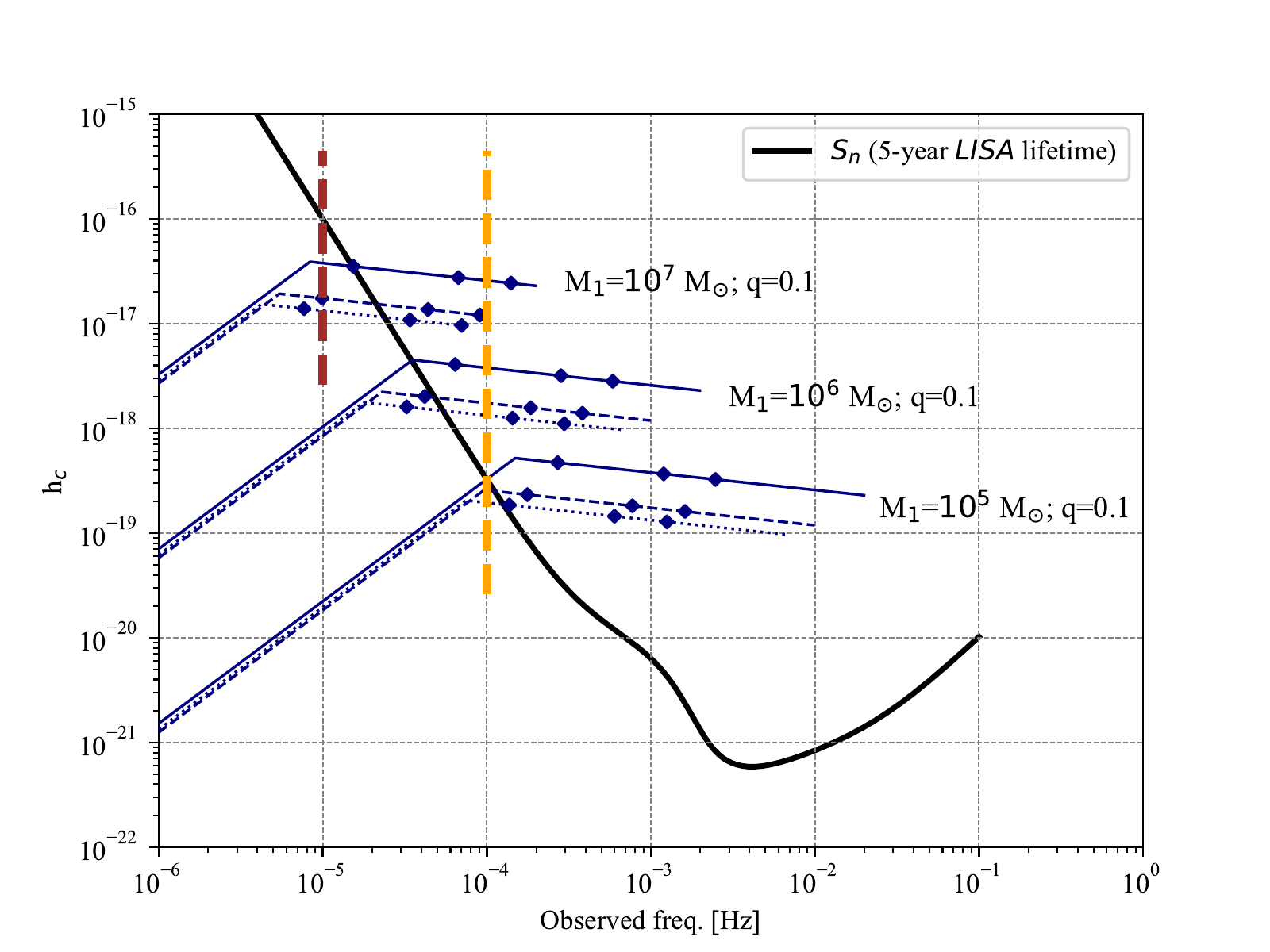}
    \caption{Characteristic GW strain of MBH binaries (blue lines) for three different primary BH masses as labelled, assuming a binary mass ratio $q=0.1$, and shown in each case for three different redshifts, $z=1$ (solid), $z=3$ (dashed) and $z=5$ (dotted). The diamonds from left to right indicate when the binary is observed to be $t_m=1$ year, 1 week
    and 1 day away from merger. The black curve shows {\it LISA}'s nominal sensitivity from \citealt{Robson2019}, assuming a mission duration of 5 years. 
    The red and orange vertical dashed lines at $10^{-5}$ and $10^{-4}$ Hz mark the possible range of a ``frequency wall" below which {\it LISA}'s sensitivity is expected to degrade significantly.}
    \label{fig:lisa_sensitivity}
\end{figure}

For illustration, we reproduce {\it LISA}'s sensitivity curve in Figure~\ref{fig:lisa_sensitivity}. The vertical red and orange dashed lines mark the possible range of the ``frequency wall", below which {\it LISA}'s sensitivity is expected to degrade rapidly. We also show illustrative examples of the evolution of MBH binaries assuming a 5--year {\it LISA} mission lifetime.
The tracks of binaries with primary masses of M$_1=10^5$M$_{\odot}$, $10^6$M$_{\odot}$ and $10^7$M$_{\odot}$ and redshift $z=1$, 3 and 5 are plotted, all with a mass ratio of $q\equiv M_2/M_1=0.1$. The characteristic strain $h_c$ is given by $\sqrt{n}h$, where $n$ is the number of cycles spent at each frequency, and $h$ is the GW strain averaged over sky location and polarisations~\citep[see, e.g.][]{Sesana2005a},
\begin{equation}
    h=\frac{8\pi^{2/3}}{10^{1/2}}\frac{G^{5/3}{M_c}^{5/3}}{c^4 r(z)} {f_r}^{2/3}.
\end{equation}
Here $M_c=[q/(1+q)^2]^{3/5} M_{\rm bh}$ is the chirp mass, $f_r=(1+z)f_{\rm obs}$ is the rest--frame GW frequency, and $f_{\rm obs}=2/P$ is the observed GW frequency, with $P$ the binary's (observed) period.

We assume for simplicity that the orbit of the MBHB is circular (as expected in the {\it LISA} band; \citealt{Munoz+2019,Zrake+2021}) and that it decays due to GW emission, which should be a good assumption at the very small separations considered here~\citep[e.g.][]{Haiman2009a}.  The rate of change in rest-frame frequency can be expressed as,
\begin{equation} \label{eq:f_dot}
    \dot{f_r} = \frac{96\pi^{8/3}G^{5/3}}{5c^5}{M_c}^{5/3}{f_r}^{11/3},
\end{equation}
where $G$ is Newton's constant and $c$ is the speed of light. The time to merger in the binary's rest-frame is
$t_{\rm m,r}$ = (3/8)$f_{\rm r}$/$\dot{f_{\rm r}}$ and the observed time to merger is $t_{\rm m} = (1+z) t_{\rm m,r}$.

Binaries with $t_m\gsim 5$ yr accumulate signal-to-noise ratio at each frequency for 
the full assumed {\it LISA} mission duration of 5~years, while binaries closer to merger spend less than this time at each frequency. This results in the ``knees" in Figure~\ref{fig:lisa_sensitivity} at the observed frequencies that correspond to $t_{\rm m}\approx 5$ yr. 
The diamonds from left to right mark the  frequencies at which the MBHB is observed to be 1 year, 1 week and 1 day to merger, and the evolution of each track is terminated at the frequency of the innermost stable circular orbit (ISCO). These examples assume binary BH mass ratios of $q=0.1$ for illustration.

As mentioned above, we will require LSST-identified MBH binaries to enter the {\it LISA} band 5-15 years after their discovery.  In general, the {\it LISA} sensitivity curve constrains the range of masses $M_{\rm bh}$, mass ratios $q$, redshifts $z$ and orbital periods $P$ of the LSST candidates that satisfy this criterion.  Given the uncertainties in the time at which the binaries may be identified in LSST and in the timing of the launch and duration of the {\it LISA} mission, for simplicity we instead require that the time to merger is $t_m=5-15$ years.  
As Figure~\ref{fig:lisa_sensitivity} shows, most binaries with masses detectable by LSST are also well above {\it LISA}'s sensitivity curve just prior to their merger. The exceptions are the most massive BHs ($\gsim 10^8~{\rm M_\odot}$) at high redshifts ($z\gsim 5$) which may merge at frequencies below {\it LISA}'s frequency wall.  These binaries are excluded from our counts below, but they constitute a very small fraction of the LSST quasar sample to begin with.

\section{Results} \label{sec:results}
In this section, we present our results for the total number of quasars that LSST can discover in its multi-year survey (\S~\ref{sec:LSST-quasar-counts}), and the ultra-short period subset of these quasars that may correspond to compact MBH binaries and will subsequently be detectable by {\it LISA} (\S~\ref{sec:LISA-quasar-counts}).

\subsection{Quasars number counts in LSST}
\label{sec:LSST-quasar-counts}

Our predictions of the total number of quasars in LSST, described in \S~\ref{sec:QLF} and \S~\ref{sec:N_tot} are shown in Figure~\ref{fig:N-z-coarse}.   In total, we find approximately $19-100$ million quasars down to the magnitude limits of $m_i=24-26$. The number counts we find
are generally somewhat higher than those predicted by the original QLF in K19.  The difference from K19 at most redshifts and magnitudes are nevertheless small, and arise only from our smoothly interpolating the best-fit double-power-law parameters, which we consider more physical.  The large majority of quasars are at faint magnitudes $m_i\gsim 24$, where our counts agree with K19 nearly exactly.
As another point of reference, in Figure~\ref{fig:N-z-coarse} we also show the number counts forecast by the LSST, and quoted in Table 10.2 in Version 2.0 of their Science Book~\cite{LSSTScienceCollaboration2009}.
As the figure shows,  the number counts we predict at the faint end are significantly above those in the LSST Science Book at these faint magnitudes.  

More specifically, down to $m_i=24$, which corresponds to the magnitude limit in a single-night LSST exposure (consisting of two separate visits), we predict that LSST will discover 19 million quasars, compared with $\sim8.4$ million in \citet{LSSTScienceCollaboration2009}. 
LSST will predominantly observe MBHs of low masses ($\sim 10^{5-6} M_{\odot}$) due to the steepness of the faint--end slope of the quasar LFs. For this reason, we extrapolate the LFs to fainter magnitudes, beyond $m_i=24$ in Figure~\ref{fig:N-z-coarse}, where no data is gathered in K19. By co-adding data from many visits during its full multi-year survey, LSST might discover objects as faint as $m_i=26$.
By integrating the QLFs between redshift 0.5 to 6.5 and $16\leq m_i\leq 26$,  we find $100$ million quasars compared to $16.7$ million in the \citealt{LSSTScienceCollaboration2009}.
Our numbers are $\sim$ 6 times higher because we adopted a QLF with a steeper faint-end slope, and because we included BH's with masses down to $\sim 10^5 {\rm M_\odot}$ (see Fig.~\ref{fig:Norb_5yr} and detailed discussion of this discrepancy in \S~\ref{sec:lf-uncertainty} below). Beyond $z\sim6$, the bandpass $\mathcal{K}$-correction is not available to convert absolute magnitude to apparent magnitude. So in order to evaluate the number of quasars at high redshift, using eq.~\ref{eq:V_eff}, we extrapolate the $\mathcal{K}$-correction with a straight line to cover $6<z<7.5$. A very small fraction of the total number is found at high redshift -- we predict approximately 38,000, 10,000 and 3,300 quasars down to $m_i$ = 26, 25 and 24, respectively, in this redshift range $6<z<7.5$.

\subsection{The number of ``verification binaries" detectable by LISA} 
\label{sec:LISA-quasar-counts}
As explained in the previous section, we hypothesise that a fraction $f(t_{\rm m})$ of the quasars $\mathcal{N}$ in the LSST catalogue are powered by ultra-compact, inspiralling MBH binaries which will enter the {\it LISA} frequency band and merge within the time $t_{\rm m}$ after their discovery, with $t_{\rm m}$ between 5 and 15 years.
This fraction is extremely small, since we have to "catch" a quasar during the last $\sim 5-15$ years of its life. The fraction can simply be approximated by the ratio $f(t_{\rm m})=t_{\rm m}$ / $t_{\rm Q}$, so that if quasars have average lifetimes of $t_{\rm Q} = 10^7$ years, and we require a MBH binary system to be within 10 years of its merger, i.e. $t_{\rm m}$ = 10 years, this fraction is $10^{-6}$.

More generally, the number of {\it LISA}-detectable quasars, $N_{\rm LISA}=[t_{\rm m}$ / $t_{\rm Q}] f_{\rm Q,b} \mathcal{N}$ simply scales linearly both 
with this ratio $(t_{\rm m}$ / $t_{\rm Q})$, and with the assumed fraction $f_{\rm Q,b}$ of quasars associated with binary MBH mergers to begin with, and can be trivially adjusted for any other assumed quasar lifetime, different time-to-merger of the MBH binary, or overall binary fraction.

We compute the characteristic strains ($h_c$) of the hypothetical binaries in the ($M_{\rm bh},z$) plane, where we assume that the binaries have a $t_m=$5 yr (observer frame) time to merger. 
We then compare their characteristic strains to the {\it LISA} sensitivity curve, shown in Figure~\ref{fig:lisa_sensitivity}. 
We find that the evolutionary track $h_c$ of 99\% of the objects eventually crosses the sensitivity curve and the $10^{-5}$ Hz ``frequency wall" and reaches the ISCO frequency above it.
The only exceptions are the most massive binaries with  $M_{\rm bh}\gsim {\rm few}\times 10^{7-9} {\rm M_\odot}$ (with the threshold mass depending on the adopted location of the frequency wall), which merge at frequencies below {\it LISA}'s sensitivity band.  However, these massive BHs comprise $\lsim 1\%$ of all LSST quasars.

\begin{figure}
    \centering
    \includegraphics[width=\columnwidth]{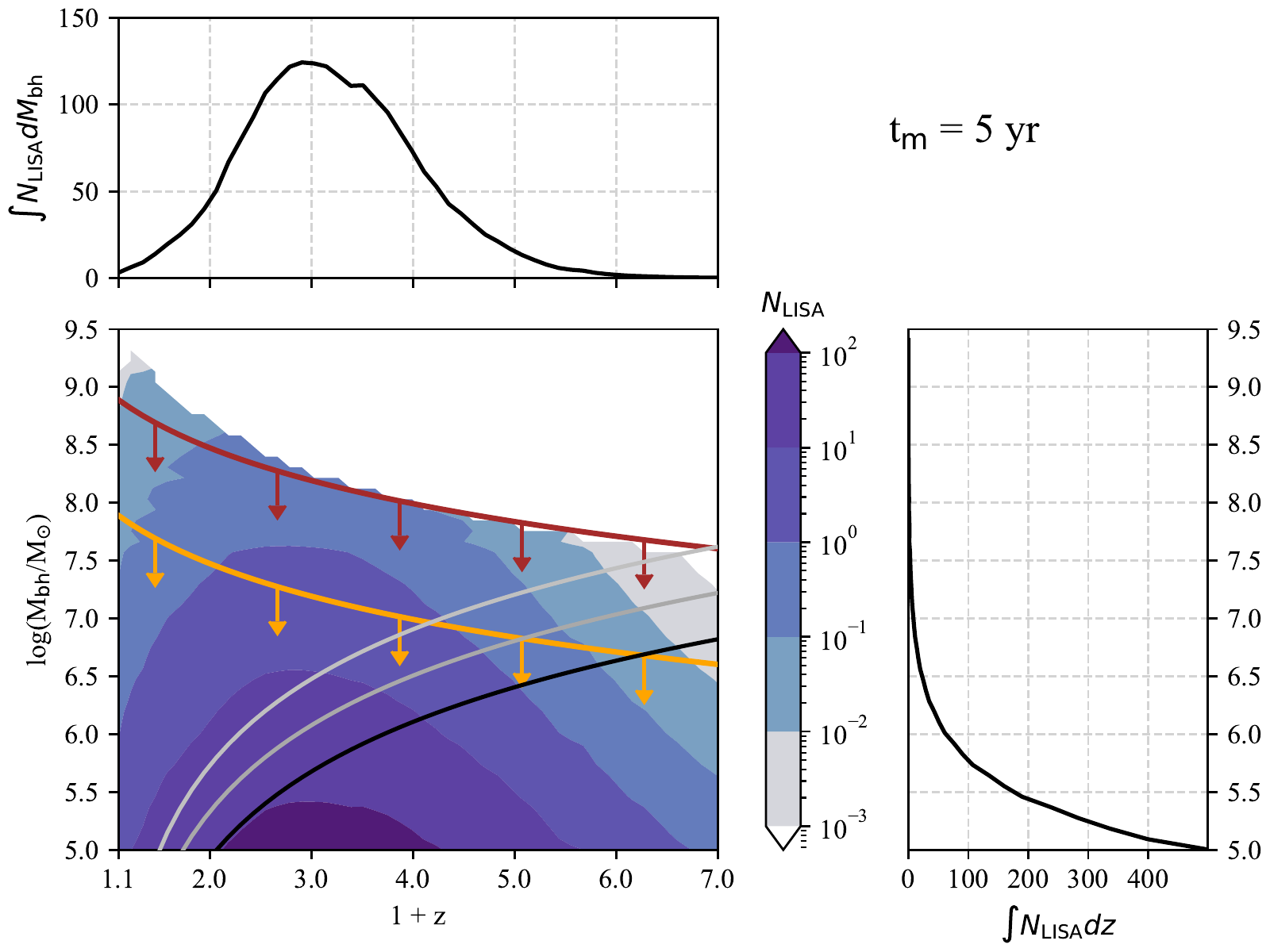}
    \caption{Number of MBH binary candidates identified as periodic quasars in LSST, which can subsequently be detected by {\it LISA}.  The contours are in units of per log$M_{\rm bh}$ per $z$.
    The top panel shows $N_{\rm LISA}$ integrated over all BH masses,   
    while the right panel shows $N_{\rm LISA}$ summed over all redshifts.
    Imposing "frequency walls" for {\it LISA} at f$_{\rm obs}>10^{-4}$ Hz (orange) or $>10^{-5}$ Hz (red) translates to upper limits on the MBH binary masses shown by the lines with downward arrows. The light grey, medium grey and black curves correspond to LSST detection thresholds of $m_i=24, 25$, and 26, respectively.
     This model assumes an observed time to merger of $t_m=5$ years and a typical quasar lifetime of $t_Q=10^7$ years. The predicted numbers scale linearly with $t_m/t_Q$.}
    \label{fig:Norb_5yr}
\end{figure}

We show $N_{\rm LISA}=N_{\rm LISA}(M_{\rm BH},z)$, 
under the assumption of $t_{\rm m}=5$ yr and $t_{\rm Q}=10^7$ yr, 
in Figure~\ref{fig:Norb_5yr} in the $(M_{\rm BH},z)$ plane.
Most objects concentrate at $z\approx 2$ at the low-mass end with $M_{\rm bh}={\rm few} \times 10^5 M_{\odot}$, which corresponds to the faint end of the QLF ($m=26$). This is unsurprising, given the steep faint-end slope of the QLF. In this figure, we also show the histograms of $N_{\rm LISA}$ integrated over all masses in the top panel, and that integrated over all redshifts in the right panel.  We observe a peak of $N_{\rm LISA}$ around $z$=2, which is consistent with the peak of the luminosity function
($\phi_{\star}$) in Figure~\ref{fig:LF-compare}. We find that {\it LISA} will detect a total of 50 quasars from the full LSST catalogue, down to $m=26$.
 
Unfortunately, not all 50 of these objects can be detected, due to further constraints. First, the possible ``frequency wall" of {\it LISA} could impose upper limits on the mass of the binary, above which binaries merge at too low frequencies to be detected.
We mark the mass limits induced by the range of a possible frequency wall in Figure~\ref{fig:Norb_5yr}. As the wall moves from 10$^{-5}$ Hz (red) to 10$^{-4}$ Hz (orange), lower-mass binaries become detectable. However, this has only a minor influence on the total number, because the majority of {\it LISA}-detectable sources concentrate at the low-mass region.

Additionally, one should take into account the uncertainty on how faint LSST would be able to observe quasars and identify their periodicity. 
The duration of observation required to identify periodic quasars will generally vary from one object to the other, depending on magnitude and period, as well as the nature of the stochastic red noise for each individual quasar. 
The nominal single-night detection threshold for quasars is $m_i\approx 24$~\citep{LSSTScienceCollaboration2009}. This corresponds to the mass limit shown by the light grey curve in Figure~\ref{fig:Norb_5yr}. Only objects with masses above this line can be detected by LSST in a single night. This cut would eliminate the large majority of fainter, lower-mass systems ($M_{\rm bh}\lsim 10^6~{\rm M_\odot}$ at $z\approx 2$), leaving 19.4 million sources in total, of which $\sim$10 are detectable by ${\it LISA}$.
However, this cut is overly conservative, because as data is accumulated over its full multi-year survey, LSST will be able to reach fainter magnitude limits.  As a rough estimate, we consider $m_i=26$ as this deeper magnitude limit, shown by the black line in Figure~\ref{fig:Norb_5yr}, and for reference, the medium grey line marks an intermediate threshold $m_i=25$.   
We report the total number of quasars left after imposing combinations of these  constraints in Table~\ref{Table:Norb_pairs}, and further discuss the issue of co-adding LSST data from multiple visits of the same field in \S~\ref{sec:folding-lc} below. 

\begin{table}
    \centering
    \begin{tabular}{c|ccc}
    & 26 mag & 25 mag & 24 mag \\ \hline
    10$^{-5}$ Hz & 50 & 23 & 10 \\
    10$^{-4}$ Hz & 48 & 21 & 8 \\
    \end{tabular}
\caption{The total number $N_{\it LISA}$ of quasars detectable by {\it LISA} from among the LSST quasar sample, for different values of the effective magnitude threshold ($m_i=24-26$; top row)
and the location of a {\it LISA} frequency wall ($10^{-4}$ or $10^{-5}$ Hz; left column). The numbers require $t_m\leq 5$ years to merger, and assume a typical quasar lifetime of $t_Q=10^7$ years, and scale linearly with $t_m/t_Q$.}
    \label{Table:Norb_pairs}
\end{table}

\begin{figure}
    \centering
    \includegraphics[width=1.1\columnwidth]{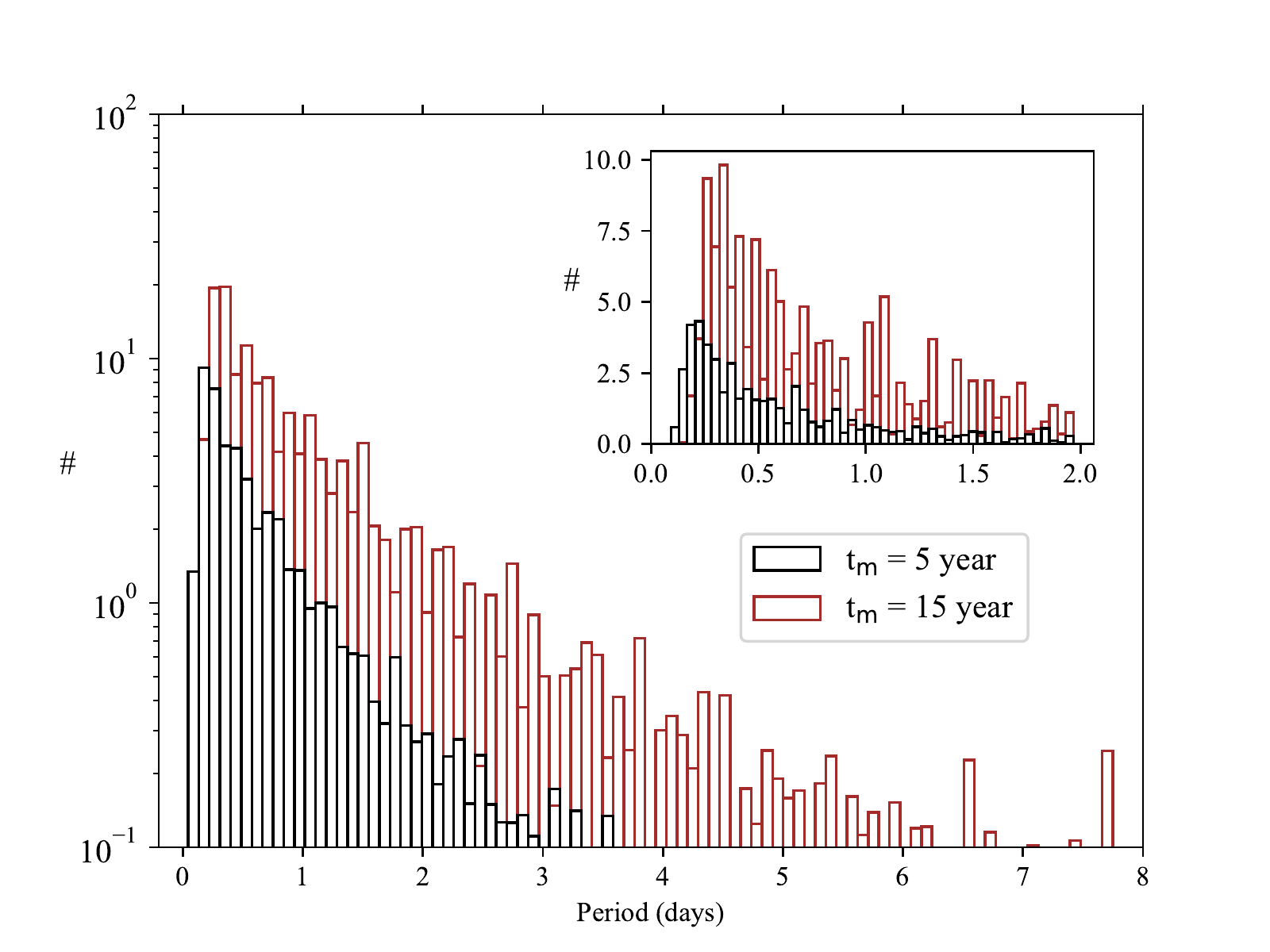}
    \caption{The orbital periods of MBH binaries $t_m=5$ years (black) and $t_m=15$ years (maroon)  prior to merger. The peaks in both distribution around 1 day indicate that most of these {\it LISA}-detectable objects have short periods of $\lsim 1-2$ days. We show the zoomed--in distributions for this range in the inset. }
    \label{fig:torb}
\end{figure}

We should also emphasise that low-mass MBH binaries with short orbital periods of $\lsim 3$ days dominate the {\it LISA}-detectable sample.
The distributions of orbital periods of binaries $t_m=5$ (black) and 15 (maroon) years prior to merger are shown in Figure~\ref{fig:torb}. We zoom in around the peaks of the two distributions in the inset figure, where the periods are shorter than 2 days. 

\begin{figure*}
    \centering
    \includegraphics[width=\columnwidth]{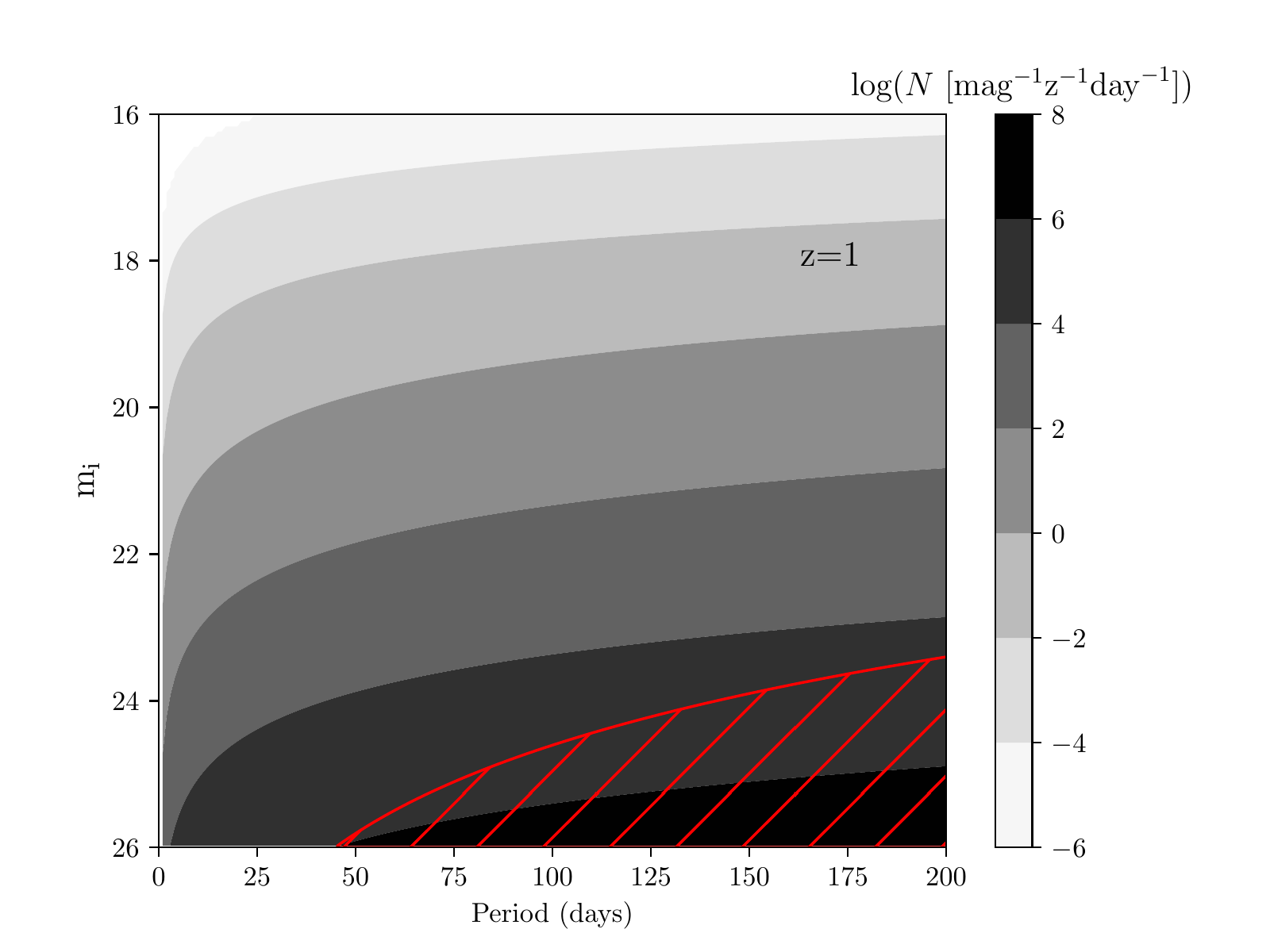}
    \includegraphics[width=\columnwidth]{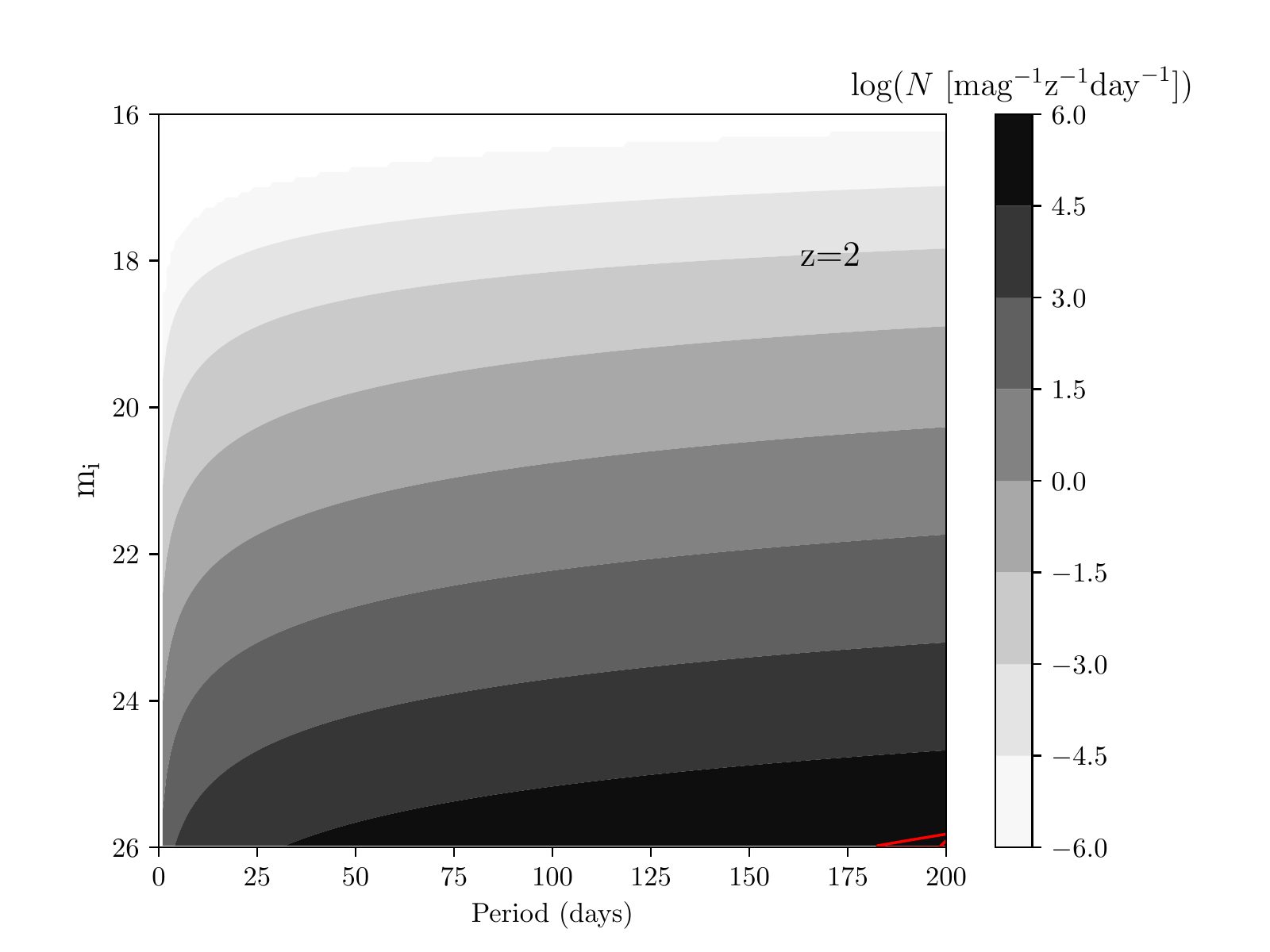}
    \hfill
    \includegraphics[width=\columnwidth]{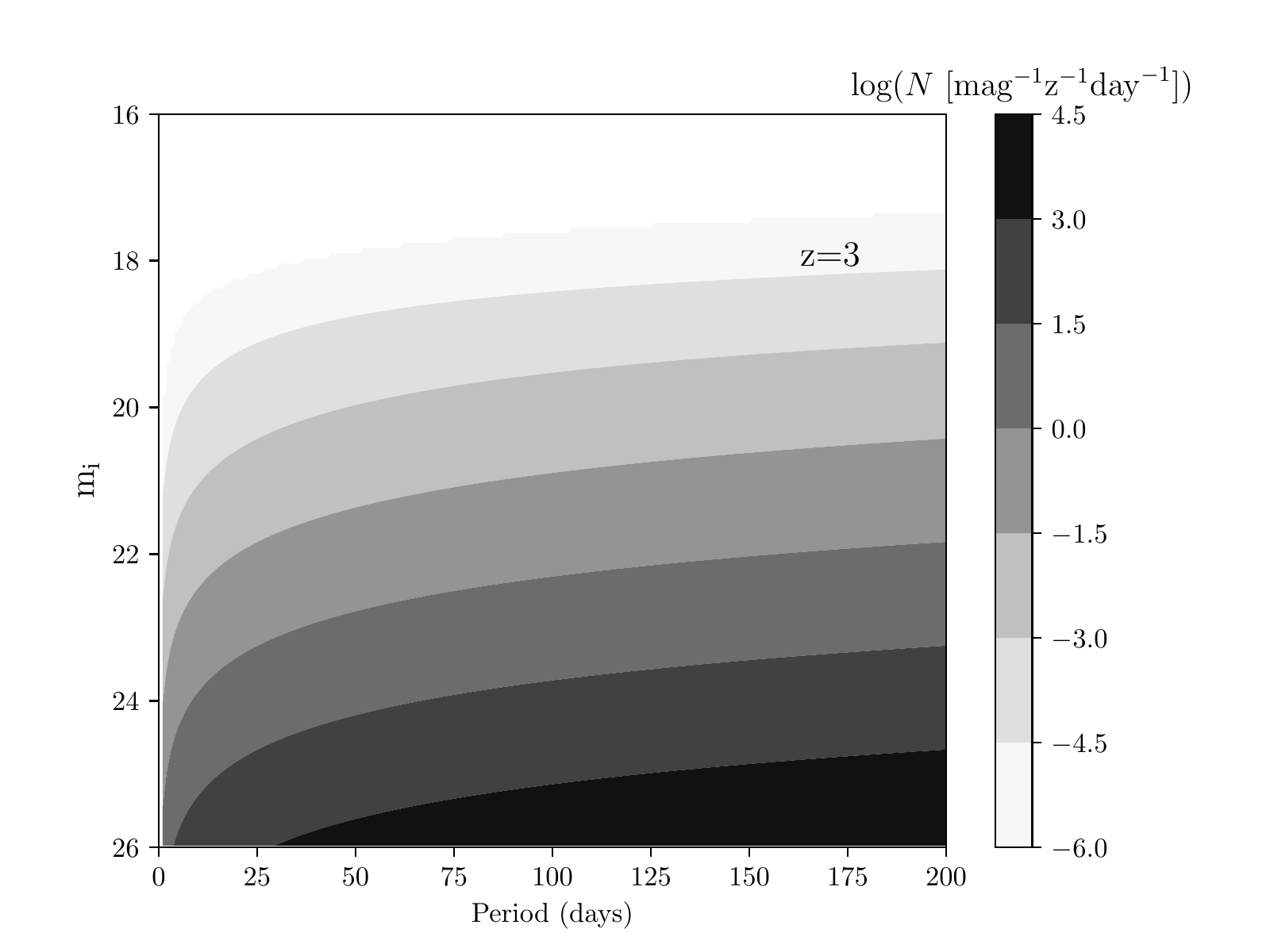}
    \includegraphics[width=\columnwidth]{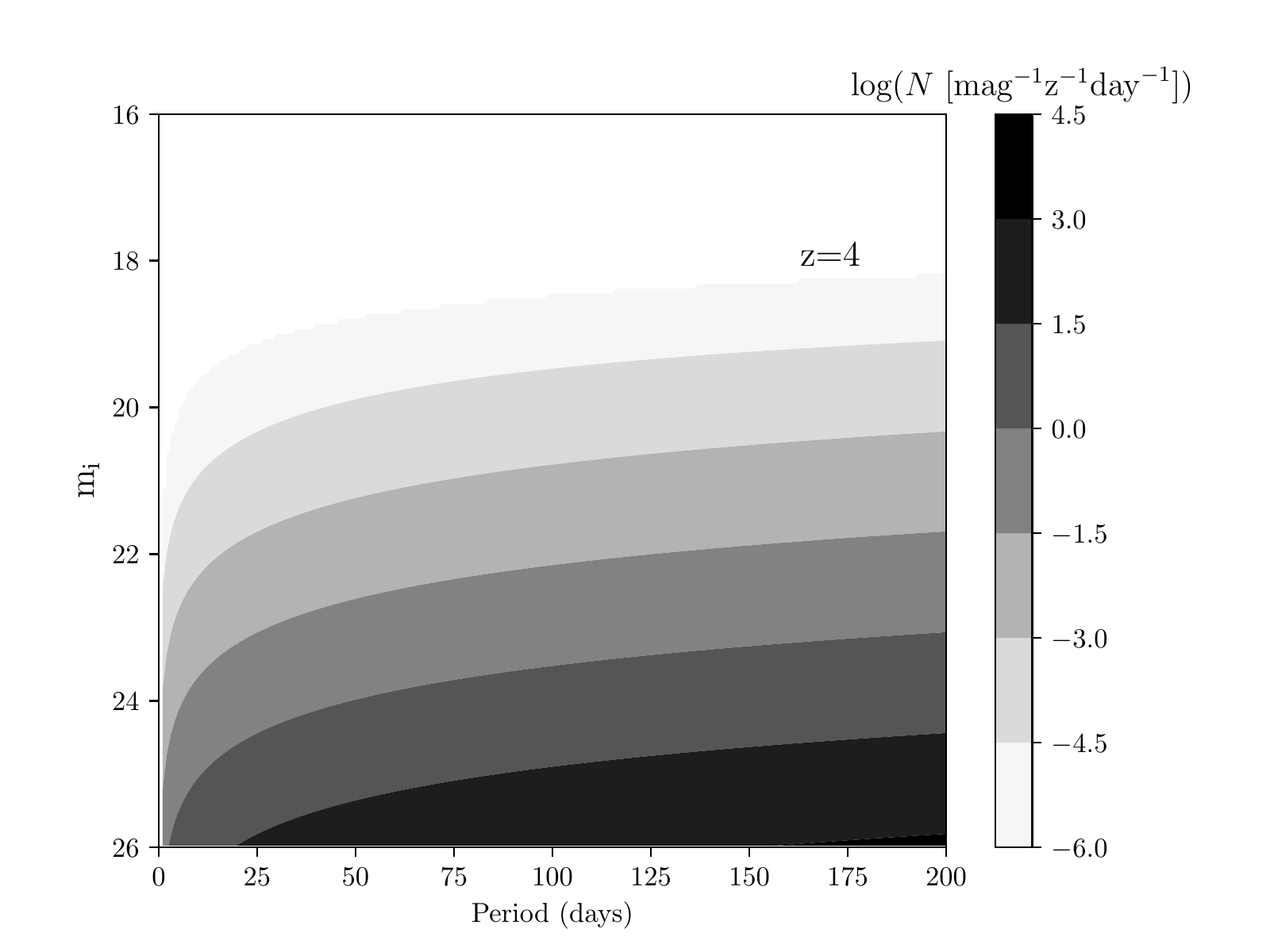}
    \caption{Expected number of quasars, $N$ (log-scale), as a function of apparent magnitude $m_i$ and observed orbital period $P$ at four different redshifts $z = 1, 2, 3$ and 4. The contours are in units of per magnitude per redshift per day, and the period-distribution assumes the binaries are inspiralling due to GWs. Binary quasars in the lower right at $z=1$ and $z=2$ (shown in hatched red) have GW inspiral times longer than the fiducial quasar lifetime of $t_Q=10^7$ years and do not exist in our simple population model.}
    \label{fig:P_contour}
\end{figure*}

The period-distributions are also of more general interest, going beyond the short-period binaries detectable by {\it LISA}.  Under our assumptions, quasars associated with binaries have lifetimes up to $t_Q$, and therefore their merger time $t_m$ can be range between $0\leq t_m \leq t_Q$.  Assuming further that the binaries inspiral purely due to GW emission, this yields a maximum orbital period ($P_{\rm max}$) defined by $t_m=(3/8)(1+z)f_r/\dot{f_r}$ as $t_m\rightarrow t_Q$, and $\dot{f_r}$ is given by eq.~\ref{eq:f_dot}.  Note that the observed period is $P=2(1+z)/f_r$, where $f_r$ is the rest-frame GW frequency. We show the number of quasars expected as a function of apparent magnitude $m_i$ and observed orbital periods  $0<P<200$ days in Figure~\ref{fig:P_contour}, at four redshifts, $z = 1,2,3$ and 4, and assuming $t_Q=10^7$ years. Note that at low redshifts, the bottom right region (marked in hatched red) corresponds to periods $P>P_{\rm max}$, where binary quasars do not exist in our simplistic model.   More generally, the period-distributions can be altered at large binary separations/periods, where the circumbinary gas can dominate the orbital evolution over GW emission (at periods as low as a $\sim$ week for the faintest quasars whose BH masses are $\sim10^5~{\rm M_\odot}$; see, e.g., Fig.1 in \citealt{Haiman2009a}).

\section{Discussion} \label{sec:discussion}

The results presented in the previous section demonstrate that there may be 10-150 "{\it LISA} Verification Binaries" in the LSST quasar catalogues. Here the lowest (10) number is for 
a conservative magnitude limit of $m_i=24$, and adopts
$(t_m/t_Q)f_{\rm Q,b}=5\times 10^{-8}$ - this may correspond, for example, $t_m=5$ years, $t_Q=3\times 10^7$ years, and $f_{\rm Q,b}$=0.3.  
The high-end (150) is for a more aggressive magnitude limit of $m_i=26$, and adopts
$(t_m/t_Q)f_{\rm Q,b}=1.5\times 10^{-7}$, which may correspond to $t_m=15$ years, $t_Q=3\times 10^7$ years, and $f_{\rm Q,b}$=0.3.  
    
These numbers rely on several assumptions, beyond the choices of $t_m, t_Q$, and $f_{\rm Q,b}$ on which the counts just depend linearly.   Here we discuss the main assumptions and related issues.

\subsection{Uncertainties in the quasar luminosity function} 
\label{sec:lf-uncertainty}

As mentioned in the previous section, the LSST Science Book forecasts a total of 16.7 million quasars, which is $\sim 6$ times lower than the number we find down to the limiting magnitude of $m_i=26$.

The first item we note is that the LSST Science Book value includes an absolute magnitude limit of $M_i<-20$ (corresponding to a BH mass of $\gsim 5\times 10^5~{\rm M_\odot}$).  
We have chosen not to include this cut, as recent years have revealed increasing evidence for the existence of low-mass BHs in galactic nuclei (see next subsection).  To assess the importance of this cut, the grey dashed curve in  Figure~\ref{fig:N-z-coarse} shows our predicted counts but with the same $M_i<-20$ cut imposed. This reduces our predictions from 19.4-100.3 million to 18.6-78.7 million quasars, still leaving a factor of 78.7/16.7=4.7 difference.

Next, the LSST Science Book predictions used an earlier QLF determination by \citet{Hopkins2007}, so we assess how much our predictions would change if we used the same QLF. Unfortunately, a precise comparison is difficult, since \citet{Hopkins2007} gives the QLF for the bolometric luminosity $L_{\rm bol}$.  However, using \citet{Runnoe+2012} to convert $L_{1450}$ to $L_{\rm bol}$ at $z=2$, we find that our counts are reduced by a factor of 1.3, 2.2, and 3.8 at $m_i=24$, 25, and 26, respectively.  We therefore conclude that our higher number counts, compared to the LSST Science Book, can be attributed almost entirely to the different adopted QLFs.
Additional, smaller differences could result from our simplified selection function.  Note, in particular, that LSST's single-image detection efficiency incorporates selections via colours, lack of proper motion, variability, and combinations with external data, and drops below unity at $m_i=24-25$~\citep{Ivezic+2014}.

To be more specific, the faint-end slope of the \citet{Hopkins2007} LF at $z=2$, $\beta=-1.27$ (their Table 3) is much shallower than the slope  $\beta=-1.98$ in K19, and the one we adopted after interpolating between redshifts ($\beta=-1.9$).  
The faint-end slope in \citet{Hopkins2007} was determined from the 2QZ survey \citep{Croom+2004}, while K19 used data from the more recent and deeper 2SLAQ survey \citep{Croom+2009}.
As discussed in Appendix B in K19, this makes the faint-end slope somewhat steeper.  However, as shown in their Fig.~B1, K19's faint-end slope, inferred from the same data, is significantly steeper even than obtained by \citet{Croom+2009}.  This disagreement may be caused by differences in 
the $\mathcal{K}$-corrections, selection-functions, and choices of binning, to which this faint-end slope is especially sensitive, and needs further investigation.   For now, we proceeded to adopt the K19-based number counts, and simply note that there is roughly an order of magnitude uncertainty in our predictions.

Finally, we note that there remain uncertainties in K19's LF determination itself.  First, the statistical errors on the fitting parameters are subdominant compared to the other uncertainties (as can be seen in the blue shaded areas in Figure~\ref{fig:LF-compare}).
However, choices in determining these parameters and interpolating them over redshift must be made. After excluding the double power--law parameters  that are affected by systematic errors in the AGN sample (open circles in Figure~\ref{fig:parameters}; note that including these discarded data points would increase our predicted counts), one can approximate the evolution of the four fitting parameters as smooth functions of redshift. K19 introduce three models in which different approaches are taken to describe the evolution of the faint-end slope. 

Given that $\phi_{\star}$, $M_{\star}$ and $\alpha$ are fitted with quadratic, cubic and linear functions 
of redshift, respectively, the faint-end slope $\beta$ is first being modelled with a double power-law in $z$, to account for the discontinuity around $z\sim 3$.
The second model uses the same exact procedure to find the four  parameters as in the first model, but excludes the sample of highest-redshift AGNs that only have rough estimates of selection functions. In the last model, they again exclude the high-$z$ samples and approximate $\phi_{\star}$, $M_{\star}$ and $\alpha$ with quadratic, cubic and linear fittings, but $\beta$ is approximated as a linear function of $z$. 

Our analysis here is closest to this third approach.  If we had adopted the other two approaches, our analysis would yield only slightly different results, approximately by 2-3\%.

\subsection{Co-adding LSST data and identifying periodicity}
\label{sec:folding-lc}

As noted above, we employed a fiducial magnitude limit of $m_i=26$, which is 2 magnitudes deeper than LSST's single-visit detection threshold of $m_i\approx 24$.
Over a full $10$-year LSST survey, assuming each field is visited on average every 3 days, or $\sim 100$ visits per year, we could naively expect that co-adding these $\sim 1000$ visits can lower the detection threshold by a factor of $\approx \sqrt{1000}\approx 30$ in flux, or by 3.7 magnitudes.
On the other hand, co-adding every visit into a single effective combined exposure would make it impossible to detect any periodicity.  As a rough estimate for a possible compromise,  $m_i=26$ represents the magnitude limit after co-adding $\sim 30$ visits. This effectively leaves $\sim 30$ independent data-points (each a co-add of 30 visits), which may facilitate both the detection of a faint, sub-single-exposure quasar and the identification of its periodicity. 

Securely identifying the periodicity will require a dedicated analysis. The expected periods are 1-2 days, which is comparable but somewhat shorter than the time between individual visits. 
For precisely uniform time-sampling, these periods would fall below the Nyquist limit and could not be detected.  However, the LSST time sampling will not be uniform, with a pair of exposures during each overnight visit, and somewhat different times elapsing between successive exposures.   This should, enable, at least in principle, to fold the effective light-curves and detect 1-2-day periodicities.  Co-adding different exposures will introduce a trade-off between going deeper (which allows LSST to discover many more quasars) but having effectively fewer points sampling the light-curve (which makes it harder to detect periodicities).   One can also imagine that choices of {\it which sets of} (not just how many) individual exposures to co-add and combine into single effective data-points will also affect the efficiency of a periodicity search.  The optimal approach to such co-adding and light-curve-folding will require a detailed dedicated analysis, which we plan to perform in a follow-up publication.

\subsection{Massive black holes below $\approx 10^6~{\rm M_{\odot}}$} 
\label{sec:low-mass-bh}

As mentioned above, in our analysis we have chosen to include quasars with BH masses down to a few $\times 10^5~{\rm M_\odot}$. Excluding quasars fainter than $M_i=-20$ (corresponding to BH masses below $5\times 10^5~{\rm M_\odot}$) would decrease our fiducial predictions from 100 million to 79 million.  This assumes Eddington ratios of $f_{\rm Edd}$ in the conversion between bolometric luminosity and BH mass; adopting a lower $f_{\rm Edd}$ would imply larger $M_{\rm bh}$.

While early dynamical searches for low-mass MBHs yielded non-detections, several campaigns in recent years have shown that a high fraction (50-80\%) of small galaxies (with masses of $M_{\rm gal}\sim 10^{9-10}~{\rm M_\odot}$) harbour nuclear MBHs with masses down to at least $M_{\rm bh}\approx 10^5~{\rm M_\odot}$.  Furthermore, these BHs obey the scaling relations such as between  $M_{\rm bh}$ and $\sigma$ (the host's velocity dispersion) seen at higher masses \citep[see the recent review by][and references therein]{Greene+2020}, and these low-mass MBHs are found among luminous AGN \citep[e.g.][and references therein]{Kimbrell+2021}.   Based on these results, we included MBHs down to these lower masses.

\subsection{Quasar lifetime and relation to MBH binaries} 
\label{sec:quasar-lifetime}

Our analysis here is based on the simplified notion that quasars are often activated by galaxy mergers, and that quasars are therefore are related to MBH binaries.  As argued in the Introduction, the connection between quasars and mergers are based in part on theoretical modelling, which helps explain the evolution of the QLF over cosmic time - particularly the rise and fall in quasar activity, and its peak at $z\sim 2$~\citep[e.g.][]{KauffmannHaehnelt2000}.  There is also well-established theoretical~\citep[e.g.][]{BarnesHernquist1991} and empirical~\citep[e.g.][]{Genzel+2010} support that major mergers drive gas to galactic nuclei and that they are associated with AGN~\citep[e.g.][]{Goulding+2018}.

It also seems inevitable that both MBHs, contained in the two merging galaxies, end up rapidly shrinking to the nucleus of the new merger-remnant galaxy~\citep[e.g.][]{Begelman+1980}.   However, there remains a major uncertainty about the precise relative timing of the quasar phase and the inspiral and eventual merger of the MBH binary. The duration of the bright quasar phase is known to last $\sim 10^{6-8}$ years~\citep[e.g.][]{Martini+2004}, while the inspiral time of the binary remains poorly understood.   Simple timescale arguments, associated the inspiral timescale with a modified viscous timescale of the AGN disk itself suggest that the coalescence of a compact MBH binary in the gas-rich nucleus of a merger remnant may be comparably short~\citep{Haiman2009a}.   It is plausible to assume that the MBH merger takes place during the luminous quasar stage, and the statistics of periodic quasar candidates to date, listed in the Introduction, are consistent with this assumption.  Nevertheless, we emphasise that this remains a key assumption that is currently poorly understood, but which will be tested by the LSST quasar variability data.

Assuming that the mergers of MBH binaries coincide with luminous quasar phases allows us to forego modelling of the evolving MBH binary population (e.g. based on underlying galaxy merger trees; \citealt{Kelley+2019}), and directly predict their number from the observed quasar luminosity functions.

\subsection{Chirping of MBHBs in LSST} 
\label{sec:chirping}

Finally, an intriguing aspect of our conclusions is that there may be a handful of ultra-short period binary quasars in the LSST survey, whose period progressively shortens due to their GW-driven inspiral. Over the timescale of years, the period will change from a $\sim$day to hours.  As an example, the observed orbital period (twice the GW period, $P=2/f_{\rm obs}$) of a binary with a total mass of $3\times 10^5~{\rm M_\odot}$ at $z=2$ (a typical "Verification Binary") will change from 2.3 days at $t_m=10$ years prior to merger to 0.6 days at $t_m=5$ years.  
Whether or not LSST data is capable of inferring this "chirp" from the optical light-curve needs to be addressed by a dedicated analysis, as described in \S~\ref{sec:folding-lc} above, but incorporating the expected rate of period-change $\dot{P}$ due to the GW emission.   A secure detection of $\dot{P}>0$, in agreement with the expected value from the GW chirp, could provide a 
smoking-gun evidence for the presence of a MBH binary.

\vspace{\baselineskip}
\section{Conclusions}
\label{sec:conclusion}

In this paper, we computed the number of faint quasars expected to be discovered in the Vera C. Rubin Observatory's LSST, and found it to be as large as 20-100 million, down to the apparent magnitudes $m_i=24-26$ reached over the multi-year survey.  We further hypothesised that a small fraction $f(t_m)$ of these sources correspond to ultra-short period binary quasars, powered by massive BH binaries at the last stages of their coalescence, i.e. within the time $t_m$ prior to their merger.

Our model is based on the quasar LF derived from a recent compilation of data from multiple AGN surveys, and the simple ansatz that the late-stage merging fraction is given by the product $f(t_m)=(t_m/t_Q)f_{\rm Q,b}$.  This expresses the product of the probability of "catching" a quasar during the last $t_m$ years of its full lifetime $t_Q$, and the underlying fraction $f_{\rm Q,b}$ of quasars that are associated with MBH mergers.

Our main result is that the LSST quasar catalogue may contain a handful, and perhaps up to several hundred, such ultra-compact binaries, which will subsequently move into {\it LISA}'s mHz frequency band, and become detectable as a GW sources.   We dub these massive BH "verification binaries", since their presence may be known before {\it LISA} is launched -- in analogy with  a handful of known Galactic white dwarf verification binaries.

These results can be interpreted in the reverse direction, as well:  once {\it LISA} detects a population of MBH binaries, it will be possible to look back in the LSST (and similar) archival data, and retroactively search for periodicities of quasars.  While in some cases, there may be a single quasar  in the three-dimensional {\it LISA} error volume~\citep{Kocsis+2006}, in the majority of cases there will be several.  The archival search should offer a way to uniquely identify the true counterpart, i.e. as the single AGN whose past periodicity matches the expected orbital period, extrapolated back in time from the {\it LISA} detection.

Our conclusions are subject to several caveats, primarily due to the poorly constrained faint-end slope of the quasar LF, and the details of the analysis required to co-add single-visit LSST data to search for the faint ($m_i\approx 26$) and ultra-short ($P\approx$ 1-2 day) period AGN.   Future work will clarify the feasibility of the discovery of these verification binaries.

\section*{Acknowledgements}
The authors thank Philip F. Hopkins, \v{Z}eljko Ivezi\'{c}, David Kipping, Girish Kulkarni, David Schiminovich and Xuejian Shen for useful discussions.  This work was supported by NASA grant NNX15AB19G and NSF grants AST-2006176 and AST-1715661.

\section*{Data availability}
No new data were generated or analysed in support of this research.

\bibliographystyle{mnras}
\bibliography{cx} 
\bsp
\label{lastpage}
\end{document}